\documentclass[prl,aps,twocolumn,superscriptaddress,showpacs]{revtex4}
\usepackage{graphicx}

\begin{document}
\renewcommand{\thefootnote}{\fnsymbol{footnote}}

\title{First direct observation of two protons in the decay of $^{45}$Fe with a TPC}

\author{J.~Giovinazzo}\affiliation{Centre d'Etudes Nucl\'eaires de Bordeaux Gradignan -
Universit\'e Bordeaux 1 - UMR 5797 CNRS/IN2P3, Chemin du Solarium, BP 120,
F-33175 Gradignan Cedex, France}
\author{B.~Blank}\affiliation{Centre d'Etudes Nucl\'eaires de Bordeaux Gradignan -
Universit\'e Bordeaux 1 - UMR 5797 CNRS/IN2P3, Chemin du Solarium, BP 120,
F-33175 Gradignan Cedex, France}
\author{C.~Borcea\footnotemark[1]}\affiliation{Centre d'Etudes Nucl\'eaires de Bordeaux Gradignan -
Universit\'e Bordeaux 1 - UMR 5797 CNRS/IN2P3, Chemin du Solarium, BP 120,
F-33175 Gradignan Cedex, France}
\author{G.~Canchel}\affiliation{Centre d'Etudes Nucl\'eaires de Bordeaux Gradignan -
Universit\'e Bordeaux 1 - UMR 5797 CNRS/IN2P3, Chemin du Solarium, BP 120,
F-33175 Gradignan Cedex, France}
\author{C.E.~Demonchy}\affiliation{Centre d'Etudes Nucl\'eaires de Bordeaux Gradignan -
Universit\'e Bordeaux 1 - UMR 5797 CNRS/IN2P3, Chemin du Solarium, BP 120,
F-33175 Gradignan Cedex, France}
\author{F.~de Oliveira Santos}\affiliation{Grand Acc\'el\'erateur National d'Ions Lourds, 
CEA/DSM - CNRS/IN2P3, Bvd Henri Becquerel, BP 55027, F-14076 CAEN Cedex 5, France}
\author{C.~Dossat}\affiliation{DAPNIA, CEA Saclay, F-91191 Gif-sur-Yvette Cedex, France}
\author{S.~Gr\'evy}\affiliation{Grand Acc\'el\'erateur National d'Ions Lourds, 
CEA/DSM - CNRS/IN2P3, Bvd Henri Becquerel, BP 55027, F-14076 CAEN Cedex 5, France}
\author{L.~Hay\footnotemark[2]}\affiliation{Centre d'Etudes Nucl\'eaires de Bordeaux Gradignan -
Universit\'e Bordeaux 1 - UMR 5797 CNRS/IN2P3, Chemin du Solarium, BP 120,
F-33175 Gradignan Cedex, France}
\author{J.~Huikari}\affiliation{Centre d'Etudes Nucl\'eaires de Bordeaux Gradignan -
Universit\'e Bordeaux 1 - UMR 5797 CNRS/IN2P3, Chemin du Solarium, BP 120,
F-33175 Gradignan Cedex, France}
\author{S.~Leblanc}\affiliation{Centre d'Etudes Nucl\'eaires de Bordeaux Gradignan -
Universit\'e Bordeaux 1 - UMR 5797 CNRS/IN2P3, Chemin du Solarium, BP 120,
F-33175 Gradignan Cedex, France}
\author{I.~Matea}\affiliation{Centre d'Etudes Nucl\'eaires de Bordeaux Gradignan -
Universit\'e Bordeaux 1 - UMR 5797 CNRS/IN2P3, Chemin du Solarium, BP 120,
F-33175 Gradignan Cedex, France}
\author{J.-L.~Pedroza}\affiliation{Centre d'Etudes Nucl\'eaires de Bordeaux Gradignan -
Universit\'e Bordeaux 1 - UMR 5797 CNRS/IN2P3, Chemin du Solarium, BP 120,
F-33175 Gradignan Cedex, France}
\author{L.~Perrot}\affiliation{Grand Acc\'el\'erateur National d'Ions Lourds, 
CEA/DSM - CNRS/IN2P3, Bvd Henri Becquerel, BP 55027, F-14076 CAEN Cedex 5, France}
\author{J.~Pibernat}\affiliation{Centre d'Etudes Nucl\'eaires de Bordeaux Gradignan -
Universit\'e Bordeaux 1 - UMR 5797 CNRS/IN2P3, Chemin du Solarium, BP 120,
F-33175 Gradignan Cedex, France}
\author{L.~Serani}\affiliation{Centre d'Etudes Nucl\'eaires de Bordeaux Gradignan -
Universit\'e Bordeaux 1 - UMR 5797 CNRS/IN2P3, Chemin du Solarium, BP 120,
F-33175 Gradignan Cedex, France}
\author{C.~Stodel}\affiliation{Grand Acc\'el\'erateur National d'Ions Lourds, 
CEA/DSM - CNRS/IN2P3, Bvd Henri Becquerel, BP 55027, F-14076 CAEN Cedex 5, France}
\author{J.-C.~Thomas}\affiliation{Grand Acc\'el\'erateur National d'Ions Lourds, 
CEA/DSM - CNRS/IN2P3, Bvd Henri Becquerel, BP 55027, F-14076 CAEN Cedex 5, France}

\begin{abstract}

The decay of the ground-state two-proton emitter $^{45}$Fe was
studied with a time-projection chamber and the emission of two protons was
unambiguously identified. The total decay energy and the half-life measured in this work
agree with the results from previous experiments. The present result constitutes the
first direct observation of the individual protons in the two-proton decay of a
long-lived ground-state emitter. In parallel, we identified for the first time directly 
two-proton emission from $^{43}$Cr, a known
$\beta$-delayed two-proton emitter. The technique developped in the 
present work opens the way to a detailed study of the mechanism of ground-state 
as well as $\beta$-delayed two-proton radioactivity.

\end{abstract}


\date{\today}

\maketitle

\renewcommand{\footnoterule}{}
\footnotetext[1]{Permanent address: NIPNE-HH, P.O. Box MG6, Bucharest-Magurele, Romania}
\footnotetext[2]{Permanent address: Laboratoire PHLAM, B\^atiment P5 - USTL, F-59655 Villeneuve d'Ascq Cedex, France}


Since the advent of machines to produce atomic nuclei far away from the valley of nuclear stability,
$\beta$-delayed particle emission and direct one-proton emission were among the tools most efficiently
used to elucidate the structure of the atomic nucleus close to the proton drip line.
The study of these decay modes allowed one to
investigate the nuclear mass surface, to determine half-lives, to establish the sequence of single-particle
levels or to improve the modeling of astrophysical processes.

For the most proton-rich nuclei with an even number of protons, Goldanskii~\cite{goldanskii60} predicted the
occurence of a new nuclear decay mode, which he termed two-proton (2p) radioactivity. He suggested that, at the proton
drip line and due to the pairing of protons,
nuclei exist, which are bound with respect to one-proton emission, but unbound for two-proton emission.
In the early 1960's, it was not clear which nuclei would be good candidates for this new
decay mode. It is interesting that the list of possible 2p emitters proposed by Goldanskii~\cite{goldanskii61}
in the 1960's contained already $^{45}$Fe, for which ground-state two-proton radioactivity was indeed
discovered~\cite{giovinazzo02,pfuetzner02} only recently.

This new type of radioactivity was observed in projectile-fragmentation experiments at the LISE3 separator of GANIL~\cite{giovinazzo02}
and at the FRS of GSI~\cite{pfuetzner02}. These experiments allowed for an unambiguous identification of this
decay mode by measuring the decay energy, which nicely fits modern decay $Q$ value predictions~\cite{brown91,ormand96,cole96},
and the half-life, which is in agreement with models linking decay energy and Coulomb and centrifugal barrier penetration
half-lives~\cite{grigorenko01,brown03}. In addition, both experiments could demonstrate that, with a very large likelihood,
there was no $\beta$ particle emission. This conclusion was achieved by a direct search for positrons~\cite{giovinazzo02} or
for the 511~keV photons from positron annihilation~\cite{pfuetzner02} as well as by the width of the 2p emission peak, which
is too narrow for a pile-up event of protons and $\beta$ particles. The most convincing piece of evidence is certainly
the observation of the daughter decay with a half-life in nice agreement with the half-life of $^{43}$Cr, the
2p daughter of $^{45}$Fe~\cite{blank05finustar}. All other possible daughters (e.g. $\beta$-delayed proton emission daughter)
can be excluded based on the experimentally measured daughter half-life.

After the observation of 2p radioactivity for $^{45}$Fe, this decay mode was also identified for
$^{54}$Zn~\cite{blank05zn54} and evidence has been found for 2p radioactivity of $^{48}$Ni~\cite{dossat05}. However, in
both experiments the decay was observed in a silicon detector, which allows only for a
determination of the total decay energy, the half-life, the branching ratio, and the absence of $\beta$ radiation, as
in the case of $^{45}$Fe. In none of these experiments, the two protons were explicitly identified. Emission of
two protons has been observed directly only from excited states (see e.g.~\cite{cable84,fynbo99,mukha06}) or from very
short lived resonances~\cite{bochkarev92,kryger95,bain96}. In the present work, we report for the first time on the direct
observation of two protons emitted by $^{45}$Fe. This is the first case of a direct two-proton observation of a
long-lived ground-state two-proton emitter.

The technique developped in the present work, the purpose of which is to determine the complete kinematics for two protons
in three dimensions, paves the way for a detailed study of ground-state
and $\beta$-delayed two-proton emission. It will enable us to investigate the decay dynamics, i.e. whether the decay
is an ordinary three-body decay or whether the two protons are correlated in energy and in angle, the sequence of 
single-particle levels beyond the drip line and more. However, this nuclear structure information can only be extracted by
a comparison to theoretical models. Therefore, a development in parallel of powerful nuclear structure and reaction models is 
essential.


The $^{45}$Fe nuclei were produced at the SISSI-ALPHA-LISE3 facility of GANIL. A primary $^{58}$Ni$^{26+}$ beam with an energy of
75~MeV/nucleon and an average intensity of 3~$\mu$A was fragmented in a $^{nat}$Ni target (200$\mu$m) in the SISSI device.
The fragments of interest were selected by a magnetic-rigidity, an energy-loss, and a velocity analysis by means of the ALPHA
spectrometer and the LISE3 separator including an energy degrader (500$\mu$m of beryllium) in the LISE3 dispersive focal plane and
transported to the LISE3 final focal plane. Time-of-flight, energy-loss, and residual energy measurements allowed for an
event-by-event identification of individual fragments. Details of the identification procedure can be found in
reference~\cite{dossat07}.

\begin{figure}[hht]
\begin{center}
\includegraphics[scale=.5]{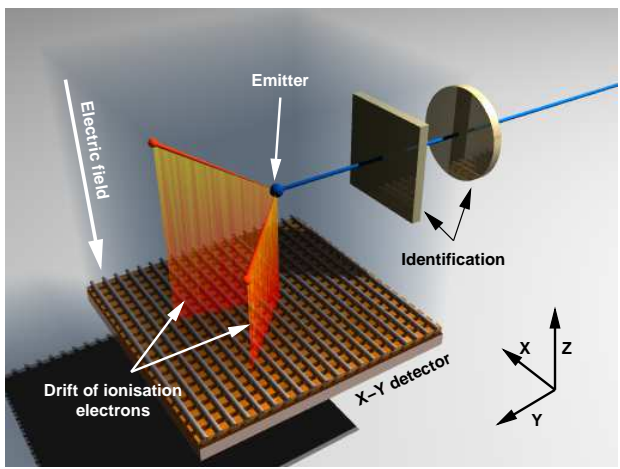}
\caption[]{Schematic representation of the time-projection chamber developed at the CEN Bordeaux-Gradignan. The nuclei
           of interest traverse first two silicon detectors (one standard detector with a thickness of 150$\mu$m and a second
           position-sensitive detector (150$\mu$m) with resistive readout) and are then
           implanted in the active volume of a gas-filled chamber, where they decay by emission of charged particles.
           The charge cloud created by the energy loss of either the heavy ions or the decay products drifts in the electric field
           of the chamber towards a set of four gas electron multipliers (GEMs, not shown), where the charges are multiplied.
           The electric potential finally directs the charges onto a two-dimensional detection plane, where two orthogonal sets of
           768 strips (pitch of 200$\mu$m) allow for a measurement of the charges deposited on each strip. Every second strip
           is equipped with an ASIC readout yielding signal height and time stamp. The other half of the strips is connected in
           groups of 64 strips to standard preamplifiers and shapers.}
\label{fig:tpc}
\end{center}
\end{figure}

Fragments transmitted to the LISE3 final focal plane were finally implanted in the center of newly developed
time-projection chamber (TPC)~\cite{blank07tpc}, which allows for a visualisation in three dimensions of implantation
and decay events (see figure~\ref{fig:tpc}). For this purpose, the charges created by the primary ionising particles
(protons or heavy ions) in the gas (Argon (90\%) - Methane (10\%)) drift in an electric field towards a set of four gas
electron multipliers (GEMs)~\cite{sauli97}, where the charges are multiplied, and are finally detected by two
orthogonal sets of 768 strips with a pitch of 200$\mu$m. The front-end electronics, based on ASIC technology,  
allows for the measurement of the charge deposited on each strip and of the arrival time of the
charge on this strip with respect to a start triggered by the first strip that fires. The ASICs were readout by 
VME multiplexed ADCs. This readout mode yielded a dead time per event of about 1.3~ms. 

The signals from different strips were gain-matched first by
injecting a pulser signal into the last GEM thus creating an image charge on each single strip. This gain matching was
refined by means of primary and fragment beams traversing the whole chamber with sufficient energy so that the energy loss
can be assumed constant over the chamber. The strips parallel to the beam direction were gain-matched by rotating the
chamber by 90$^0$. The TPC was extensively tested off-line with $\alpha$ sources and on-line with different beams, which
consituted a proof of the working principle of the chamber. Details of this new instrument will be published 
elsewhere~\cite{blank07tpc}.

After a setting on $^{52}$Ni, a well known $\beta$-delayed proton emitter with decay energies around 1.2~MeV (see e.g.~\cite{dossat07}), the main
setting of the SISSI-ALPHA-LISE3 facility was optimised for the production and selection of $^{45}$Fe. In addition to
$^{45}$Fe, a large number of $^{43}$Cr nuclei were also produced and implanted in the TPC in the same setting.
$^{43}$Cr is a $\beta$-delayed two-proton ($\beta$2p) emitter, for which, however, the two protons were never directly
observed~\cite{borrel92,giovinazzo01a,dossat07}.


In the present paper, we will present only the results obtained from the energy signals of the strips, which allows for an
unambiguous identification of the two protons emitted by $^{45}$Fe. The analysis of the time signals, which will give rise to a
three-dimensional reconstruction of the proton tracks, is much more complicated and is beyond the scope of the present paper. 
The complete analysis and the final results of the present experiment will be published in a subsequent paper.

\begin{figure}[hbb]
\begin{center}
\includegraphics[scale=0.43]{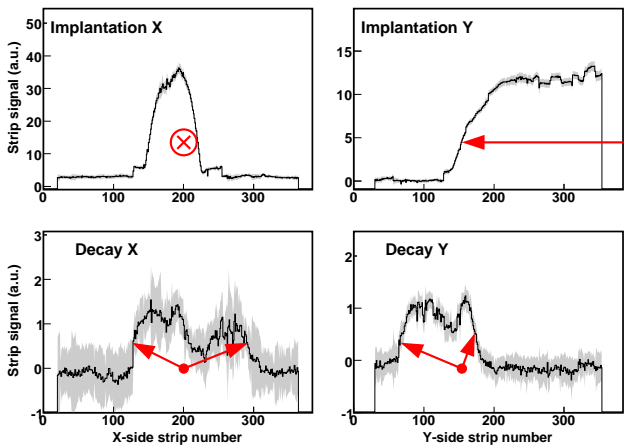}
\caption[]{Upper part: The figure shows a typical implantation event for $^{45}$Fe. The ion enters parallel to the X strips (left figure) 
           and stops at a certain depth in Y direction (right figure). This plot allows to determine the stopping position of the ion, where then
           the decay has to occur. Lower part: The decay of $^{45}$Fe takes place by two-proton emission. The decay tracks start,
           within the position resolution of about 6-7~mm (FWHM), at the stopping point of the implantation event described above.
           The arrows are indicative only of the tracks of the two protons. A horizontal arrow means that the proton propagates
           perpendicular to the strips, whereas a vertical arrow indicates that the proton track is parallel to the strips. 
           The shaded zone corresponds to the error bars due to the gain matching procedure.}
\label{fig:fe45_implant}
\end{center}
\vspace*{-0.5cm}
\end{figure}

In order to accept a pair of implantation and decay events, the decay tracks in X and Y have to start where the implantation tracks
end. Figure~\ref{fig:fe45_implant} shows an implantation event for $^{45}$Fe. The beam enters parallel
to the X strips and stops at a certain Y depth in the active volume of the TPC (see figure~\ref{fig:fe45_implant}).
There, the decay tracks of the two protons emitted have to start.

In the present experiment, we observed a total of 10 $^{45}$Fe implantations, which could be correlated with
decays. About 5 other decay events are lost to a large extent due to the data acquisition dead time mentioned above
and the short half-life of $^{45}$Fe ($T_{1/2}$~= 1.75$^{+0.49}_{-0.28}$ms)~\cite{dossat05}.
In these cases, the treatment of the $^{45}$Fe implantation event prevents the registration of its decay.

\begin{figure}[hht]
\begin{center}
\includegraphics[scale=0.42]{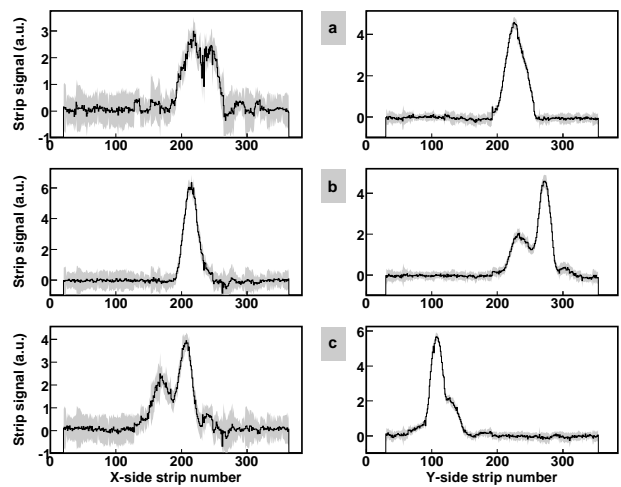}
\caption[]{The figure shows three of the ten decay events observed after implantation of $^{45}$Fe. We have chosen
           here events, where the two protons can be nicely identifed. For two of the ten events,
           the two protons can not be separated, as they fire the same strips.}
\label{fig:fe45_events}
\end{center}
\vspace*{-0.2cm}
\end{figure}

Four $^{45}$Fe decay events are shown in the lower part of figure~\ref{fig:fe45_implant} and in figure~\ref{fig:fe45_events}. 
We will shortly explain the event in figure~\ref{fig:fe45_implant} in the following.
On the lower left panel (X strips), one sees the starting point of the two protons around channel 200, in agreement with the
maximum of the implantation event in figure~\ref{fig:fe45_implant}, and traces extending on the left and the right of this point, 
testimony of the two protons being emitted across a number of strips. The lower right panel indicates a starting point around 
strip 160. One observes a large bump on which is superimposed a narrow one. The first indicates that one proton flies more or 
less perpendicular to the Y strips, whereas the second proton has a trajectory paralell to these strips. 
Therefore, this figure clearly evidences the presence of two charged particles in the decay of $^{45}$Fe. 
There is no other explanation possible for such a decay pattern.

This is the first direct observation of the emission of two protons in the ground-state decay of a long-lived two-proton emitter.
As mentioned above, all previous evidence, even if equally conclusive, was indirect.
Although the third coordinate Z is not yet determined - it will follow from the drift-time analysis, one can already state
at this stage that a variety of decay configurations is observed and not only one type, e.g. a back-to-back emission. 

\begin{figure}[hht]
\begin{center}
\includegraphics[scale=0.36]{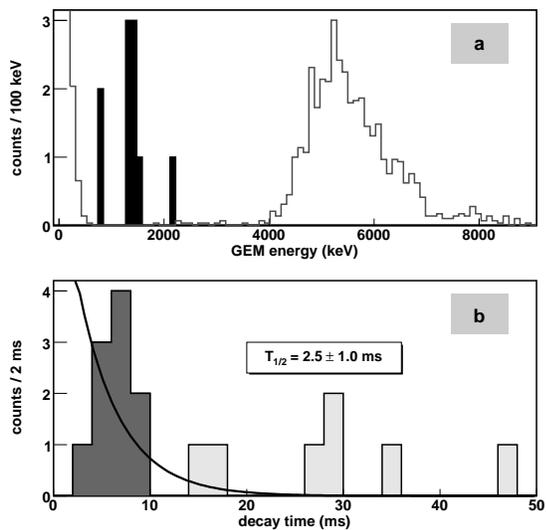}
\vspace*{-0.2cm}
\caption[]{(a) Total decay energy of the $^{45}$Fe events as determined with the last GEM. The low-energy events (in black) show the $^{45}$Fe
           decay, whereas the high-energy events are from a triple-$\alpha$ source. The energy of the $^{45}$Fe events correspond
           roughly to the expected decay energy of 1.15~MeV~\cite{dossat05}. (b) Half-life of $^{45}$Fe as determined from
           the time difference of the implantation events and the decay events (black events). The half-life is (2.5$\pm$1.0)~ms.
           The events in gray are the daughter-decay events which were not included in the fit.}
\label{fig:fe45_decay}
\end{center}
\vspace*{-0.5cm}
\end{figure}

After background subtraction, the charges deposited by decay events on the different strips can be integrated and are
proportional to the energy of the decay event. Charges on X and Y strips were integrated separately. In addition,
the different GEMs yield also the total decay energy. Therefore, the charge signal from the GEMs is sent to a
charge-sensitive pre-amplifier and a shaper. In figure~\ref{fig:fe45_decay}a, we plot this charge signal from the last GEM.
A rough energy calibration was achieved by means of a triple $\alpha$ source. The decay energy thus determined corresponds
roughly to the expected decay $Q$ value of 1.15~MeV~\cite{dossat05}.

The time difference between an implantation event and its subsequent decay event allows for a determination of the half-life
of $^{45}$Fe. This spectrum is shown in figure~\ref{fig:fe45_decay}b. A fit with an exponential, excluding the first 2~ms
which are affected by the data acquisition dead time, yields a half-life value of
(2.5$\pm$1.0)~ms. This value is nice agreement with the average value from previous experiments of
1.75$^{+0.49}_{-0.28}$~ms~\cite{dossat05}.

\begin{figure}[hht]
\begin{center}
\includegraphics[scale=0.4]{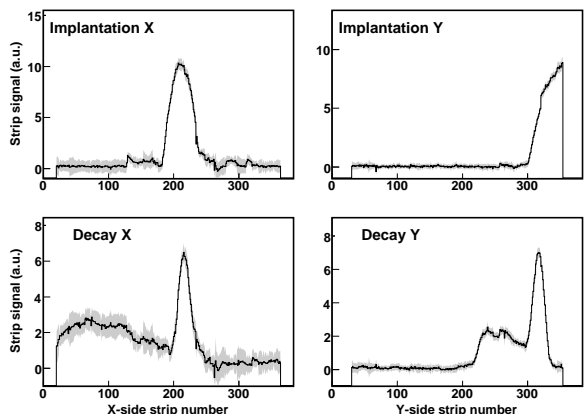}
\vspace*{-0.6cm}
\caption[]{Upper part: Implantation event of a $^{43}$Cr nucleus in the TPC. The nucleus is identified by means of energy-loss
           and time-of-flight measurements. Lower part: Decay of $^{43}$Cr as observed with the TPC. The traces of the two
           protons can be seen in the X and Y direction.}
\label{fig:cr43}
\end{center}
\end{figure}

As mentioned above, the setting of LISE3 on $^{45}$Fe allowed us also to study the decay of $^{43}$Cr, a known $\beta$-delayed
two-proton emitter, for which the $\beta$2p decay was only identified by means of the total decay
energy~\cite{borrel92,giovinazzo01a,dossat07}. The present experiment
with the TPC enabled us for the first time also to visualise the emission of two protons from this nucleus. Although the active
volume of the present version of the TPC is too small to stop all the protons in this decay due to their much higher energy
(up to about 2~MeV for each proton), the two protons are nonetheless clearly observable for most of the decay events.
One of these decay events together with its preceding implantation is shown in figure~\ref{fig:cr43}. The two-proton tracks
are visible in both directions.

A detailed study of this decay and the search for a possible angular correlation between the two protons can be envisaged
either by increasing the active volume of the TPC or by working at higher pressure. In this way, many other $\beta$2p
emitters could be identified and studied.


In summary, we have performed the first experiment, which demonstrates unambiguously that $^{45}$Fe emits indeed two protons
in most of its decays. This is the first time that ground-state two-proton emission of a long-lived (longer than 10$^{-21}$s)
isotope is observed directly. Thus, ground-state two-proton radioactivity is definitively established as a nuclear decay
mode.

A more detailed analysis will try to determine how the two protons share the energy and thus to elucidate the decay
mechanism which governs two-proton radioactivity. The complete analysis of the energy and of the time signals should
allow for a more detailed access to the decay kinematics. This will enable us to determine the proton-proton angle
which then can be compared to model predictions for the decay pattern of two-proton radioactivity. However, to
establish a detailed picture of the decay process, higher-statistics data (50 - 100 implantation and decay events) are needed, 
which can be obtained in a future experiment.

\section*{Acknowledgement}

We would like to thank the whole GANIL and, in particular, the LISE staff  and the DAQ group for their support during
the experiment. This work was partly funded by the Conseil r\'egional d'Aquitaine.


\begin{thebibliography}{10}

\bibitem{goldanskii60}
V.~I. Goldanskii, Nucl. Phys. {\bf 19},  482  (1960).

\bibitem{goldanskii61}
V.~I. Goldanskii, Nucl. Phys. {\bf 27},  648  (1961).

\bibitem{giovinazzo02}
J. Giovinazzo {\it et~al.}, Phys. Rev. Lett. {\bf 89},  102501  (2002).

\bibitem{pfuetzner02}
M. Pf{\"u}tzner {\it et~al.}, Eur. Phys. J. {\bf A14},  279  (2002).

\bibitem{brown91}
B.~A. Brown, Phys. Rev. C {\bf 43},  R1513  (1991).

\bibitem{ormand96}
W.~E. Ormand, Phys. Rev. C {\bf 53},  214  (1996).

\bibitem{cole96}
B.~J. Cole, Phys. Rev. C {\bf 54},  1240  (1996).

\bibitem{grigorenko01}
L. Grigorenko {\it et~al.}, Phys. Rev. C {\bf 64},  054001  (2001).

\bibitem{brown03}
B.~A. Brown and F.~C. Barker, Phys. Rev. C {\bf 67},  041304  (2003).

\bibitem{blank05finustar}
B. Blank, AIP Conf. Proc. {\bf 831},  352  (2005).

\bibitem{blank05zn54}
B. Blank {\it et~al.}, Phys. Rev. Lett. {\bf 94},  232501  (2005).

\bibitem{dossat05}
C. Dossat {\it et~al.}, Phys. Rev. C {\bf 72},  054315  (2005).

\bibitem{cable84}
M.~D. Cable {\it et~al.}, Phys. Rev. C {\bf 30},  1276  (1984).

\bibitem{fynbo99}
H. Fynbo {\it et~al.}, Phys. Rev. C {\bf 59},  2275  (1999).

\bibitem{mukha06}
I. Mukha {\it et~al.}, Nature {\bf 439},  298  (2006).

\bibitem{bochkarev92}
O.~V. Bochkarev {\it et~al.}, Sov. J. Nucl. Phys. {\bf 55},  955  (1992).

\bibitem{kryger95}
R.~A. Kryger {\it et~al.}, Phys. Rev. Lett. {\bf 74},  860  (1995).

\bibitem{bain96}
C. Bain {\it et~al.}, Phys. Lett. {\bf B373},  35  (1996).

\bibitem{dossat07}
C. Dossat {\it et~al.}, submitted to Nucl. Phys. A  (2007).

\bibitem{blank07tpc}
B. Blank {\it et~al.}, to be published  .

\bibitem{sauli97}
F. Sauli, Nucl. Instr. Meth. A {\bf 386},  531  (1997).

\bibitem{giovinazzo01a}
J. Giovinazzo {\it et~al.}, Eur. Phys. J. {\bf A11},  247  (2001).

\bibitem{borrel92}
V. Borrel {\it et~al.}, Z. Phys. A {\bf 344},  135  (1992).

\end{thebibliography}

\end{document}